\documentclass[journal,twoside]{IEEEtran}
\usepackage{cite}
\usepackage{amsmath,amssymb,amsfonts}
\usepackage{algorithmic}
\usepackage{graphicx}
\usepackage{textcomp}
\usepackage{array}
\usepackage{booktabs,color,epsfig,hyperref,url}
\usepackage{subfigure}
\usepackage{algorithm}

\def\BibTeX{{\rm B\kern-.05em{\sc i\kern-.025em b}\kern-.08em
    T\kern-.1667em\lower.7ex\hbox{E}\kern-.125emX}}
\begin{document}
\title{Fisher Matrix Based Fault Detection for PMUs Data in Power Grids }
\author{Ke Chen, Dandan Jiang, Bo Wang and Hongxia Wang
\thanks{D. Jiang was supported by Key technologies for coordination and interoperation of power distribution service resource, Grant No. 2021YFB2401300; NSFC Grant No. 11971371 and  Natural Science Foundation of Shaanxi Province, Grant No.  2020JM-049.}
\thanks{K. Chen and D. Jiang are with
the School of Mathematics and Statistics, Xi'an Jiaotong University, Xi'an, China (e-mail: kechencke@163.com; jiangdd@xjtu.edu.cn).}
\thanks{B. Wang (corresponding author) and H. Wang are with
the School of Electrical Engineering and Automation, Wuhan University, Wuhan, China (e-mail: whwdwb@whu.edu.cn; 2018282070092@whu.edu.cn).}}
\maketitle

\begin{abstract}
Abnormal event detection is critical in the safe operation of power system.
In this paper, using the data collected from phasor measurement units (PMUs), two methods based on Fisher random matrix are proposed to detect faults in power grids. Firstly, the fault detection matrix is constructed and the event detection problem is reformatted as a two-sample covariance matrices test problem. Secondly, the central limit theorem for the linear spectral statistic of the Fisher matrix is derived and a test statistic for testing faults is proposed.
To save computing resources, the screening step of fault interval based on the test statistic is designed to check the existence of faults. Then two point-by-point methods are proposed to determine the time of the fault in the interval. One method detects faults by checking whether the largest sample eigenvalue falls outside the supporting set of limiting spectral distribution of the standard Fisher matrix, which can detect the faults with higher accuracy. The other method tests the faults based on the statistic proposed, which has a faster detection speed.
Compared with existing works, the simulation results illustrate that two methods proposed in this paper cost less computational time and provide a higher degree of accuracy.
\end{abstract}


\section{Introduction}
\label{sec:introduction}
\IEEEPARstart{T}{his} paper is driven by the need of event detection of power systems using a stream of big data from phasor measurement units (PMUs). Because of the characteristics of high temporal resolution \cite{1}, precise time stamping \cite{2}, and high dimensionality \cite{1,3}, PMUs provides big data rich in power system state information. Therefore, it is worth considering how to make full use of the massive data resources from PMUs to do fault detection in the distribution and transmission networks \cite{4,5}.

The easy access to big data from PMUs contributes to the development of data-driven disturbance detection methods in power systems \cite{6}.
For example, support vector machine (SVM) is used to classify the events to evaluate the power quality (PQ) in \cite{7,8}, where different features are extracted from the big data collected, and then fed as input vector for event identification.
With the use of time series, \cite{9,10} predict fault in power systems with the long short-term memory (LSTM). And \cite{11} discriminates the frequency disturbances caused by various generation and transmission events using a deep convolutional neural network (CNN).
In spite of the good ability to extract complex relations among the input, the low generalization ability of above training-testing approaches makes them difficult to adapt to real power systems\cite{12}, as the real disturbance data are difficult to get, which makes the models above difficult to learn the fault characteristics.

Another data-driven method in fault detection is the statistical one. For instance, based on the principal components analysis (PCA), \cite{13} and \cite{14} detect faults in power systems by reconstructing the state evaluation matrix using the principal components in subspace, which help to reduce the computation but inevitably lose some information contained.
What's more, the unsuitable number of samples or the PCA window may fail this method, because there exists a ``strict'' limitation for the dimension-over-samples ratio \cite{15,16}. Furthermore, the fault can be distinguished using random matrix theory (RMT), due to the theoretically independently identically distribution (i.i.d.) characteristics of big data from PMUs when the power grid is in a normal state.
For instance, \cite{17} proposes a single-phase grounding fault detection method based on Mar${\rm\check{c}}$enko-Pastur (M-P) law, and \cite{18} identifies voltage change and locates the event according to spectral distribution change. However, the above RMT based methods have relatively lower accuracy because they focus on the correlation within the data and have large random errors, resulting in a high misjudgment rate.

To address the problems discussed above, based on the large-dimensional Fisher matrix in RMT, this paper proposes two fault-detection methods for power systems.
Firstly, the fault detection of power grids is reformatted as a two-sample covariance matrices test.
The test statistic is constructed according to the difference between the two population covariance matrices, and the limiting distribution of test statistic is derived by theoretical tools in RMT. Then the shrinking range of faults is screened out by the proposed hypothesis testing method.
Two point-by-point detection methods based on the sliding time window are proposed to detect faults within each shrinking interval of faults.
One method calculates the largest eigenvalue of the Fisher matrix and determines the faults by checking whether it falls outside the supporting set of the limiting spectral distribution (LSD) of the standard Fisher matrix. The other method determines the faults by rejecting or accepting the null hypothesis based on the statistic proposed.
The contributions of this paper are summarized as follows:

1) The power grid state is mapped to the matrix constructed, then based on Fisher matrix, the differences between the covariance matrices of normal and abnormal states are used to derive the fault detection indicator. In addition, the application of Fisher matrix free of population to fault detection of power grids can avoid the errors brought by the misuse of model.

2) To cut the computation, the interval of faults is shrunk through the screening method based on hypothesis testing, which enables two proposed methods to improve the speed of fault detection.

3) Two fast detection methods are proposed to detect the faults in power grids, combining the interval screening method with the largest-eigenvalue-based and statistic-based methods, respectively. The two methods can be used in distribution and transmission networks, and improve the speed of fault detection in large samples and dimensions situations. The first and second methods are suitable for scenarios requiring high accuracy and high computational speed, respectively.

The rest of this paper is arranged as follows. In Section \uppercase\expandafter{\romannumeral2}, we introduce the central limit theorem (CLT) to construct a hypothesis testing method for fault detection and determine the existence interval of fault by this method. In Section \uppercase\expandafter{\romannumeral3}, two point-by-point detection methods are applied in each shrinking interval. One is based on the largest eigenvalue of Fisher matrix, and the other is based on the hypothesis testing method proposed in Section \uppercase\expandafter{\romannumeral2}. In Section \uppercase\expandafter{\romannumeral4}, the effectiveness of the proposed methods is validated by simulation. Conclusions
are in Section \uppercase\expandafter{\romannumeral5}.

\section{Testing Faults by CLT for Large-Dimensional Fisher Matrix\label{sec:2}}

\subsection{Construction of Fault-detection Matrix\label{sec:2.1}}
The state evaluation matrix is constructed by the data from a wide-area measurement system.
The data of $p$ buses equipped with synchronous phasor measurements are collected as samples. The vector $\boldsymbol{x}_t=(x_{1t},\dots,x_{pt})'$ is obtained at sampling time $t$.
As the sampling time increases, $T$ vectors  $\boldsymbol{x}_1,\dots,\boldsymbol{x}_T$ are collected, and the state evaluation matrix $\boldsymbol{X}$ is constructed by all the time vectors as follows:
\begin{equation}\label{state evaluation matrix}
	\boldsymbol{X}=
	\begin{pmatrix}
		x_{11}&x_{12}&\cdots&x_{1T}\\
		x_{21}&x_{22}&\cdots&x_{2T}\\
		\vdots&\vdots&\ddots&\vdots\\
		x_{p1}&x_{p2}&\cdots&x_{pT}\\
	\end{pmatrix}.
\end{equation}
Each row of $\boldsymbol{X}$ is for the same measuring point, and each column is for the same sampling time.
The normalization of $\boldsymbol{X}$ is $\widetilde{\boldsymbol{X}}$, which consists of the following elements:
\begin{equation}\nonumber
	\tilde{x}_{ij}=\frac{x_{ij}-\mu_i}{\sigma_i},i=1,\dots,p; j=1,\dots,T,
\end{equation}
where $\mu_i=\sum_{j=1}^T x_{ij}/T$ and $\sigma_i^2=\sum_{j=1}^T (x_{ij}-\mu_i)^2/(T-1)$. So we get $\tilde{x}_{ij}$ with mean 0 and variance 1. For convenience, hereafter this text, $\widetilde{\boldsymbol{X}}$ is used to represent the normalization of $\boldsymbol{X}$, columns of $\widetilde{\boldsymbol{X}}$ are $\tilde{\boldsymbol{x}}_t=(\tilde{x}_{1t},\dots,\tilde{x}_{pt})', t=1,\dots,T$.

\subsection{Fault Detection based on Hypothesis Testing Method\label{sec:2.2}}
The data obtained from PMUs follow the different population distributions when the power system operates in a normal state and an abnormal state, respectively.
Suppose that the power system operates abnormally at sampling time $\tau$.
The state evaluation matrix $\boldsymbol{X}=\left( \boldsymbol{x}_1,\dots,\boldsymbol{x}_T\right) $ is segmented into $\boldsymbol{X}^{(1)}=\left(\boldsymbol{x}_{1},\dots,\boldsymbol{x}_\tau \right)$ and $\boldsymbol{X}^{(2)}=\left(\boldsymbol{x}_{\tau+1},\dots,\boldsymbol{x}_{T}\right)$. The distribution of their populations is different, which means there exists a change point at sampling time $\tau$.
Thus from the view of the data, the fault detection of power systems is equivalent to change-point detection of the state evaluation matrix.

Therefore, the change-point model is first formulated as follows. For the state evaluation matrix $\boldsymbol{X}=\left( \boldsymbol{x}_1,\dots,\boldsymbol{x}_T\right) $, the hypothesis is expressed as
\begin{equation}
\mathcal{H}_0 : \boldsymbol{x}_t \sim F_0, \quad t=1,\dots,T\label{HO}
\end{equation}
against the alternative
\[
\begin{aligned}
\mathcal{H}_1 : \boldsymbol{x}_t \sim F_0, \quad t=1,\dots,\tau  \quad \\
\text{and} \quad  \boldsymbol{x}_t \sim F_1, \quad t=\tau+1,\dots,T,
\end{aligned}
\]
where $F_0$ and $F_1$ are  the probability distribution functions from two different populations.

Suppose $\boldsymbol{\Sigma}_1$ and $\boldsymbol{\Sigma}_2$ are the population covariance matrices for two different populations, respectively. The normalization of $\boldsymbol{X}^{(1)}$ and $\boldsymbol{X}^{(2)}$ are $\widetilde{\boldsymbol{X}}^{(1)}$ and $\widetilde{\boldsymbol{X}}^{(2)}$, which can be seen as samples from the two different populations with covariance matrices $\boldsymbol{\Sigma}_1$ and $\boldsymbol{\Sigma}_2$.
If the system operates normally, it can be assumed that $\widetilde{\boldsymbol{X}}^{(1)}$ and $\widetilde{\boldsymbol{X}}^{(2)}$ are from the same population. They also have the same numeral characteristics, such as covariance. Once the covariance matrices of the two populations are not equal, that is, $\widetilde{\boldsymbol{X}}^{(1)}$ and $\widetilde{\boldsymbol{X}}^{(2)}$ come from different populations, the system is in abnormal state.

Therefore, we consider testing the change point in (\ref{HO}) by the following hypothesis test
$$\mathcal{H}_0:\boldsymbol{\Sigma}_1=\boldsymbol{\Sigma}_2\quad v.s. \quad \mathcal{H}_1:\boldsymbol{\Sigma}_1\neq\boldsymbol{\Sigma}_2$$
or equivalently
\begin{equation}
\mathcal{H}_0:\boldsymbol{\Sigma}_1\boldsymbol{\Sigma}_2^{-1}=\boldsymbol{I}\quad v.s. \quad \mathcal{H}_1:\boldsymbol{\Sigma}_1\boldsymbol{\Sigma}_2^{-1}\neq \boldsymbol{I},\label{HOinv}
\end{equation}
if  $\boldsymbol{\Sigma}_2$ is invertible.

Obviously, the above fault-detection problem is reformatted as a two-sample covariance matrices test. Within this context, much literature has focused on and explored this issue. In \cite{Cai}, \cite{Zhang}, \cite{Jing}, \cite{Zou}, different test statistics are proposed and their asymptotic distributions are derived. For example, \cite{Zou} proposes the statistic based on the difference between the null and alternative hypotheses, such as combining two measures $\mbox{tr}\left\lbrace(\boldsymbol{\Sigma}_1-\boldsymbol{\Sigma}_2)^2\right\rbrace$ and
$\mbox{tr}\left\lbrace(\boldsymbol{\Sigma}_1\boldsymbol{\Sigma}^{-1}_2- \boldsymbol{I})^2\right\rbrace$, and deriving their joint distribution.
For the convenience of application, this paper simplifies the statistic and only applies one of the measures
$\mbox{tr}\left\lbrace(\boldsymbol{\Sigma}_1\boldsymbol{\Sigma}^{-1}_2- \boldsymbol{I})^2\right\rbrace$, and its limiting distribution to construct the test statistic.
By the spectral analysis of the large-dimensional Fisher matrix in RMT, the proposed testing method can be applied to fault detection of large-dimensional data in power grids. Before that, we introduce some basic notations and preliminary knowledge of the large-dimensional Fisher matrix.

The unbiased sample covariance matrices corresponding to the $\boldsymbol{\Sigma}_1$ and $\boldsymbol{\Sigma}_2$ are  denoted as
$$\boldsymbol{S}_1=\frac{1}{\tau-1}\sum_{k=1}^{\tau}\left( \tilde{\boldsymbol{x}}_k-\bar{\boldsymbol{x}}^{(1)}\right) \left(  \tilde{\boldsymbol{x}}_k-\bar{\boldsymbol{x}}^{(1)}\right) ^\ast$$
and
$$\boldsymbol{S}_2=\frac{1}{T-\tau-1}\sum_{k=\tau+1}^{T}\left( \tilde{\boldsymbol{x}}_k-\bar{\boldsymbol{x}}^{(2)}\right) \left(  \tilde{\boldsymbol{x}}_k-\bar{\boldsymbol{x}}^{(2)}\right) ^\ast,$$
respectively, where $\bar{\boldsymbol{x}}^{(1)}=\sum_{k=1}^{\tau}\tilde{\boldsymbol{x}}_k/\tau,$ $\bar{\boldsymbol{x}}^{(2)}=\sum_{k=\tau+1}^{T}\tilde{\boldsymbol{x}}_k/(T-\tau)$, and $^\ast$ stands for complex conjugate and transpose.
We define the large-dimensional Fisher matrix as
\begin{equation}
	\boldsymbol{F}=\boldsymbol{S}_1\boldsymbol{S}_2^{-1}
	\label{Fmat}
\end{equation}
under the large-dimensional setting
\begin{equation}
\begin{aligned}
	\frac{p}{\tau-1}=y_\tau\rightarrow y_1\in (0,+\infty), \\ \frac{p}{T-\tau-1}=y_T\rightarrow y_2\in (0,1), \label{LDsetting}
\end{aligned}
\end{equation}
as $p, \tau,$ and $T$  go to infinity. 
If the null hypothesis in \eqref{HOinv} holds,  
the matrix $\boldsymbol{F}$ in \eqref{Fmat} is a so-called standard Fisher matrix, of which empirical eigenvalues have an LSD $F_{y_1,y_2}$, with a density function given as blow by \cite{Bai},
	\begin{equation}
	\begin{aligned}
		p_{y_1,y_2}(x)=
		\begin{cases}
			\frac{(1-y_2)\sqrt{(b-x)(x-a)}}{2\pi x(y_1+y_2 x)}, &if \ a\leq x\leq b;\\
			0,&otherwise.\\
		\end{cases}
	\end{aligned}\label{p(x)}
	\end{equation}
Here $a={(1-h)^2}\big /{(1-y_2)^2}$, $b={(1+h)^2}\big/{(1-y_2)^2}$, and $h=\sqrt{y_1+y_2-y_1y_2}$. Furthermore, if $y_1>1$, then $F_{y_1,y_2}$ has an additional point mass $1-1/y_1$ at the origin.

We denote ${\bf n}=(\tau, T-\tau)$ and let $F_{\bf n}(x)$ being the empirical spectral distribution (ESD) of the Fisher matrix $\boldsymbol{F}$. Then a linear spectral statistic (LSS) of the random matrix $\boldsymbol{F}$  is expressed as
\[ \int f (x)\mbox{d} F_{\bf n}(x)=
\frac{1}{p}\sum\limits_{i=1}^p f(\lambda_i), \quad f \in  \mathcal{A},
\]
where $\mathcal{A}$ is a set of analytic functions that are defined on an open region including the supporting set of the continuous part of the LSD $F_{y_1, y_2}$, $f$ is an arbitrary analytic function in the set $\mathcal{A}$, and $\lambda_i$'s are the real eigenvalues of the $p\times p$ square matrix $\boldsymbol{F}$. The empirical process is defined as
$G_{\bf n} : = \{G_{\bf n}(f)\}$ indexed by $\mathcal{A}$ ,
\begin{equation}\nonumber
G_{\bf n}(f)= p\cdot \int_{-\infty}^{+\infty} f(x)\left(F_{\bf n}-
F_{y_\tau, y_T}\right)(\mbox{d} x), \quad f \in  \mathcal{A},\label{Gdef}
\end{equation}
where
$F_{y_\tau, y_T}$ is the analogue of $F_{y_{1}, y_{2}}$, except that the parameter $y_1, y_2$ are replaced by $y_\tau, y_T$.

We define that $\kappa=2$ if all the $\boldsymbol{x}_t$ are real variables and $\kappa=1$ if they are complex.
According to \cite{Zheng}, the CLT for LSS of large-dimensional Fisher matrix is given by the following.

For the matrix $\boldsymbol{F}=\boldsymbol{S}_1\boldsymbol{S}_2^{-1}$ in (\ref{Fmat}),
if the large-dimensional setting
(\ref{LDsetting}) and the null hypothesis (\ref{HOinv}) hold, then for the analytic functions $f_1,\dots,f_k$ in the set $\mathcal{A}$,
the random vector $\left\lbrace G_{\bf n}(f_1),\dots,G_{\bf n}(f_k)\right\rbrace$ converges weakly to a Gaussian vector, whose means and covariance functions can be calculated by the contour integral formulas in Theorem~3.2 in \cite{Zheng}.

Next, let's recall the measure $\mbox{tr}\left\lbrace(\boldsymbol{\Sigma}_1\boldsymbol{\Sigma}^{-1}_2- \boldsymbol{I})^2\right\rbrace$ for the hypothesis test (\ref{HOinv}), it is reasonable to use the statistic $\mbox{tr}\left\lbrace(\boldsymbol{F}-\boldsymbol{I})^2\right\rbrace$ as the estimator of the measure.
Then the test statistic is defined as
\begin{equation}
L=\nu_g^{-1/2}\left\lbrace  \mbox{tr}\left\lbrace (\boldsymbol{F}- \boldsymbol{I})^2\right\rbrace - p F_{y_\tau, y_T}(g)- \mu_g\right\rbrace,\label{Lstat}
\end{equation}
where
$$F_{y_\tau, y_T}(g)
=\frac{y_\tau+y_T-y_\tau y_T  + y_T^2- y_T^3}{(1-y_T)^3 },$$
$$
\begin{aligned}
    \mu_g= &\frac{(\kappa-1)(2h^2 y_2 + h^2 - 2y_2^3 + 3y_2^2)}{(1-y_2)^4}
+\frac{\beta_1 y_1}{(1-y_2)^2} \\
&+ \frac{\beta_2(-2y_1 y_2^2 +2y_1 y_2- 2y_2^3 + 3y_2^2+y_2)}{(1-y_2)^3},
\end{aligned}$$
and
$$
\begin{aligned}
	\nu_g=&\frac{\kappa(2h^4 + 4h^2(h^2 - y_2^2 + 2y_2)^2)}{(1-y_2)^8}\\
	&+\frac{4(\beta_1 y_1+\beta_2 y_2)(h^2 - y_2^2+ y_2)^2}{(1-y_2)^6}
\end{aligned}
$$
are calculated according to Theorem~2.1 in \cite{Zou}, here $g(x)=(x-1)^2$.

For the null hypothesis  $\mathcal{H}_0:\boldsymbol{\Sigma}_1\boldsymbol{\Sigma}_2^{-1}=\boldsymbol{I}$, we assume that the large-dimensional setting (\ref{LDsetting}) holds, by the Theorem~3.2 in \cite{Zheng}, then the test statistic $L$ defined in (\ref{Lstat}) follows the standard Gaussian distribution under $\mathcal{H}_0$.	
A simulation to assess the performance of the limiting distribution of test statistic $L$ is given in the Supplementary Material.
The rejection region with significance level $\alpha$ is constructed as $W_\alpha=\{L:|L|\geq U_{1-\alpha/2}\}$, where $U_{1-\alpha/2}$ is the $1-\alpha/2$ th quantile of the standard Gaussian distribution.
When the null hypothesis is rejected, there exists a change point in the sampling time interval $\left[ 1, T\right] $.
This means the operation state of the power grid was abnormal during the sampling time.

\subsection{Screening Out the Interval of Faults\label{sec:2.3}}
From the above analysis, the test statistic $L$ can well judge whether faults occur in the sampling time $\left[1, T\right] $. Therefore, the range of faults can be shrunk by testing the smaller sampling time interval. This reduces the workload for the subsequent point-by-point detection of the faults.

For the state evaluation matrix $\boldsymbol{X}$ in (\ref{state evaluation matrix}),
to ensure that the covariance matrix is invertible, a suitable width $D$ slightly larger than $p$ is selected.
Now the data are segmented into $\boldsymbol{X}_i^{(1)}=\left(  \boldsymbol{x}_{t_{i-1}+1},\dots,\boldsymbol{x}_{t_i} \right)$,
$\boldsymbol{X}_i^{(2)}=\left( \boldsymbol{x}_{t_i+1},\dots,\boldsymbol{x}_{t_{i+1}}\right)$ at each sampling time $t_i$, where $t_i=iD, i=1,\dots,N$, $t_0=0$, and $t_{N+1}=T$. $N=[T/D]-1$, where $[\cdot]$ denotes the integer truncation function.
For each segmented data, the normalization of $\boldsymbol{X}_i^{(1)}$ and $\boldsymbol{X}_i^{(2)}$ are $\widetilde{\boldsymbol{X}}_i^{(1)}$ and  $\widetilde{\boldsymbol{X}}_i^{(2)}$, which are regarded as samples from two populations, and the covariance matrices for each population are defined as $\boldsymbol{\Sigma}_i^{(1)}$ and $\boldsymbol{\Sigma}_i^{(2)}$.
Faults are detected by testing the hypothesis (\ref{HOinv}) at each sampling time $t_i$:
\begin{equation}
	\mathcal{H}_{0i}:\boldsymbol{\Sigma}_i^{(1)}{\boldsymbol{\Sigma}_i^{(2)}}^{-1}=\boldsymbol{I}\quad v.s. \quad \mathcal{H}_{1i}:\boldsymbol{\Sigma}_i^{(1)}{\boldsymbol{\Sigma}_i^{(2)}}^{-1}\neq \boldsymbol{I}.\label{H0i}
\end{equation}
Let
$$
\begin{aligned}
	v_i=
	\begin{cases}
		1,& \mathcal{H}_{0i} \ is\ rejected\\
		0,&otherwise\\
	\end{cases}
\end{aligned}
$$
be the result of each test (\ref{H0i}). So we get a set of test results $\left\lbrace v_i,i=1,\dots,N\right\rbrace $.
Suppose there are $M$ sampling points to reject the null hypothesis.
In this set, $M$ intervals like $\left[t_{i-1}+1,t_{i+1} \right]$ can be obtained by finding all indexes $i$ that satisfy $v_i=1,i=1,\dots,M$. If there are intersections among these $M$ intervals, then taking the union of the intervals that produce the intersections. Thus we obtain $J$ disjoint intervals $\left[t_{l_j-1}+1,t_{m_j+1} \right] ,j=1,\dots,J$, where $1\leq l_1\leq m_1<\dots<l_J\leq m_J\leq N$. Here $M$ and $J$ are arbitrarily positive integers and $J\leq M\leq N$. Because the null hypothesis is rejected within these intervals, these intervals are the smaller intervals where faults may exist. As a result, the overall interval $\left[ 1,T\right] $ shrinks to a number of smaller intervals $\left[t_{l_j-1}+1,t_{m_j+1} \right]$.
The algorithm can be expressed as follows:

\begin{algorithm}
	\caption{Screening out the interval of faults}
	\label{alg2}
	\begin{algorithmic}[1]	
		\REQUIRE
		The state evaluation matrix $\boldsymbol{X}$; the width $D$;\\
		\ENSURE
		The intervals $\left[t_{l_j-1}+1,t_{m_j+1} \right] ,j=1,\dots,J$;
		\FOR{$i=1:N$}
		\STATE
		The normalization of $\boldsymbol{X}_i^{(\ell)}: ~\widetilde{\boldsymbol{X}}_i^{(\ell)} , \ell=1,2$;\\
		\STATE
		Calculate the sample covariance matrices $\boldsymbol{S}_i^{(1)},\boldsymbol{S}_i^{(2)}$;\\
		\STATE $\boldsymbol{F}_i=\boldsymbol{S}_i^{(1)}{\boldsymbol{S}_i^{(2)}}^{-1}$;\\
		$L_i=\nu_g^{-1/2}\left\lbrace  \mbox{tr}\left\lbrace (\boldsymbol{F}_i- \boldsymbol{I})^2\right\rbrace - p F_{y_\tau, y_T}(g)- \mu_g\right\rbrace$;\\
		\IF{$L_i \in W_\alpha$}
		\STATE $v_i=1$;
		\ENDIF
		\ENDFOR
		\STATE Find all indexes $i$ satisfy $v_i=1$ in set $\left\lbrace v_i,i=1,\dots,N\right\rbrace $;\\
		\RETURN $\left[t_{i-1}+1,t_{i+1} \right],i=1,\dots,M$;\\
		\STATE Find the intervals which have intersections and take the union of these intervals;\\
		\RETURN $\left[t_{l_j-1}+1,t_{m_j+1} \right] ,j=1,\dots,J$;\\
	\end{algorithmic}		
\end{algorithm}

By using this method, the shrinking intervals of faults can be screened out.
Within each shrinking interval $\left[t_{l_j-1}+1,t_{m_j+1} \right] $,
the two point-by-point detection methods proposed in the next section are used to determine the time of faults. Through the screening by the hypothesis testing method, only focus on the data within the shrinking interval and it can save computational work.
Since the methods below focus on only one shrinking interval at a time, the data matrix $\left(\boldsymbol{x}_{t_{l_j-1}+1},\boldsymbol{x}_{t_{m_j+1}} \right) $ for each shrinking interval replaces the state evaluation matrix $\boldsymbol{X}$ during detection.

\section{Determination of the Time of Faults\label{sec:3}}
\subsection{DELE: Determination by Largest Eigenvalue\label{sec:3.1}}
The spectral distribution of a random matrix can reflect the statistical characteristics of matrix elements, such as correlation. And it can be associated with the overall operation state of power systems, reflect the operation of systems and identify potential operational risks. This provides theoretical tools for fault detection. Reference \cite{MSR} carries out fault detection based on the ring law and mean spectral radius (MSR) in RMT; \cite{17} uses the correlation between M-P law and power systems state to detect faults of power grids. The previous methods
take the data as a whole for fault detection, which involves a large calculated amount when the sample size is large. To save computing resources and reduce computing time, the whole data matrix is segmented into two matrices with relatively small sample numbers. Based on the segmentation, this paper proposes a new method using the spectral distribution of large-dimensional Fisher matrix to detect the operation state of power systems.

When the power system is in normal operation, the data of each segment come from the same population, so $\boldsymbol{\Sigma}_1$ is equal to $\boldsymbol{\Sigma}_2$. Then the ESD of $\boldsymbol{F}$ converges to the given LSD $F_{y_1,y_2}$ in (\ref{p(x)}), which shows that the eigenvalues of $\boldsymbol{F}$ fall within the supporting set $(a,b)$. When a fault occurs, the LSD of $\boldsymbol{F}$ will change, that is, some extreme eigenvalues of $\boldsymbol{F}$ will exceed the upper boundary $b$ of supporting set.

\subsubsection{Adaptability Analysis of Fault Detection\label{sec:Adaptability}}
To validate the applicability of LSD of the Fisher matrix in detecting faults. This section uses simulated data of a distribution network with 80 nodes ($p=80$) based on PSCAD in Fig. 8 in \cite{17} for analysis. With a sampling frequency of 20,000 Hz, each branch is set to be one node, and the synchronized phasor measurement is installed at each node to obtain the zero-sequence current.

Considering two cases: normal operation and abnormal operation.
For all cases, the simulation time is 0.5s. In abnormal operation, node 66 is grounded by 10 $\Omega$, grounding fault starts at 0.25s and is cut off at 0.45s. Setting $\tau=500$, then $b=2.8163$ by calculation.

\begin{figure}[htbp]
	\centering
	\subfigure[]
	{
	    \begin{minipage}[b]{.45\linewidth}
	    	\centerline{
	  	\includegraphics[scale=0.3]{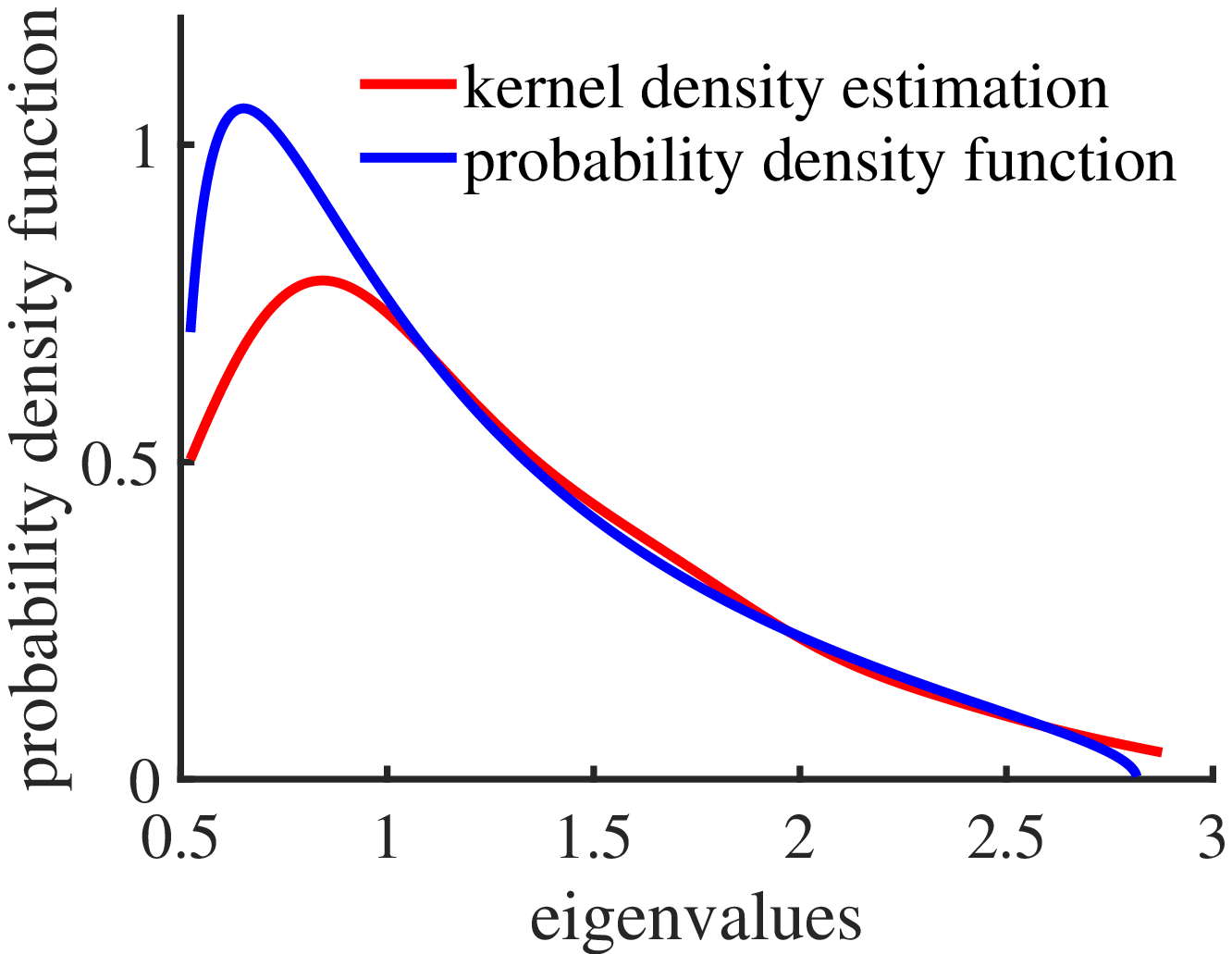}}	
	    \end{minipage}
}
	\subfigure[]
	{
		\begin{minipage}[b]{.45\linewidth}
			\centerline{
   	\includegraphics[scale=0.3]{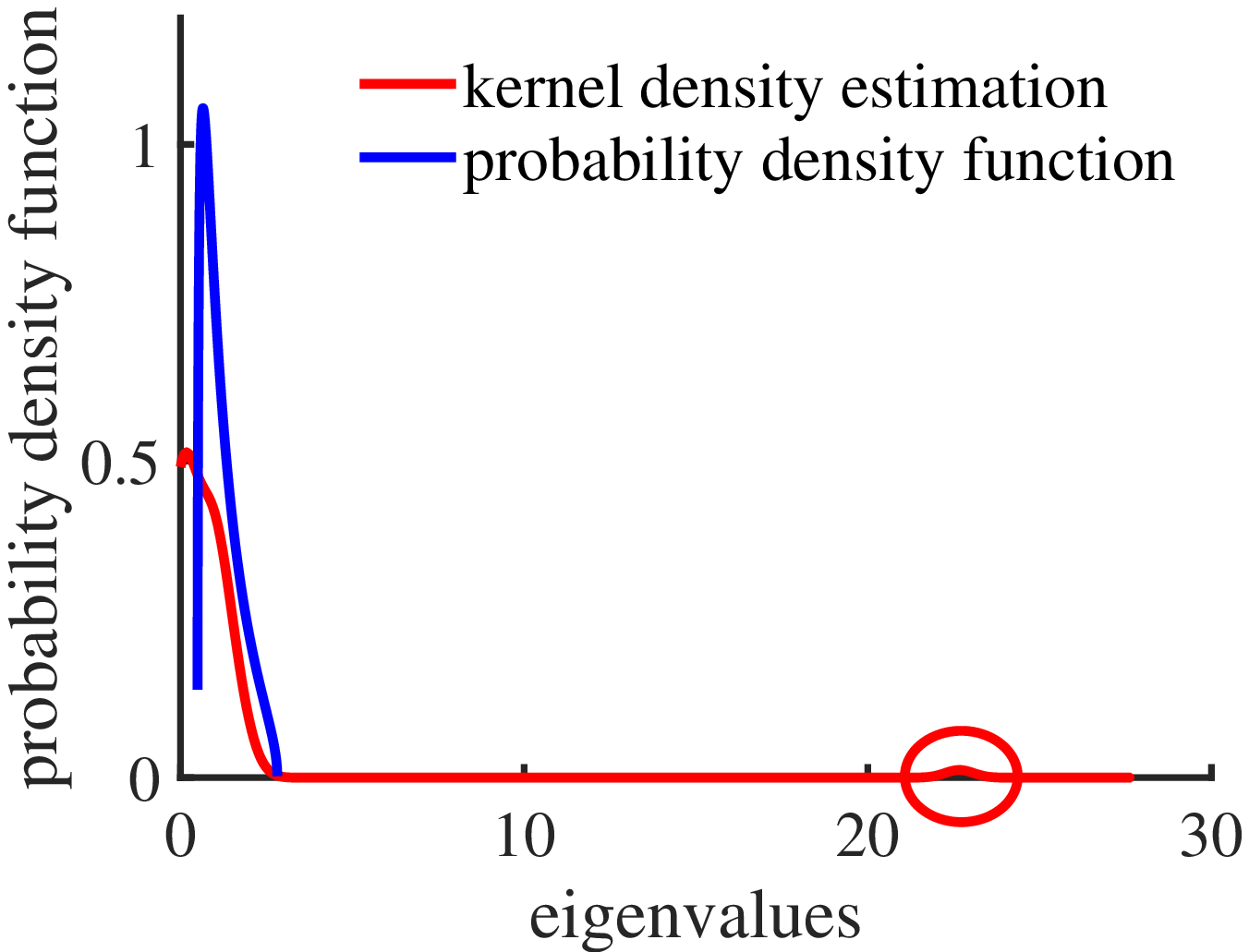}}	
		\end{minipage}
	}
\caption{Comparison between ESD of $\boldsymbol{F}$ and LSD $F_{y_1,y_2}$ under different operating conditions. (a) Normal operation. (b) Abnormal operation.}\label{fig:1}
\end{figure}

The distribution of eigenvalues of $\boldsymbol{F}$ under two cases are shown in Fig. \ref{fig:1}.
The red line is the kernel density estimation of eigenvalues of $\boldsymbol{F}$, and the blue line is the probability density function $p_{y_1,y_2}(x)$ of LSD $F_{y_1,y_2}$.

Firstly, as seen from Fig. \ref{fig:1}(a), when the power grid works normally, the kernel density is consistent with $p_{y_1,y_2}(x)$. The supporting sets of both $p_{y_1,y_2}(x)$ and kernel density are included in $(a,b)$.
Secondly, when the system works abnormally, the density function of eigenvalues of $\boldsymbol{F}$ will not be consistent with $p_{y_1,y_2}(x)$ due to the inequality of $\boldsymbol{\Sigma}_1$ and $\boldsymbol{\Sigma}_2$.
And there exists some eigenvalues of $\boldsymbol{F}$ larger than $b$ (circled in red). The supporting sets of kernel density fall outside $(a,b)$, as shown in Fig. \ref{fig:1}(b).

It is adaptable to judge whether the system is operating abnormally by checking whether the largest eigenvalue of ESD falls outside the supporting set of $F_{y_1,y_2}$.
Therefore, in the following subsection, the time of fault will be determined by comparing whether the largest eigenvalue of the Fisher matrix is larger than the upper bound $b$ of the supporting set.
The operational process of proposed method is simply conducted, avoiding the complex feature extraction of fault signals.

\subsubsection{Algorithm for DELE\label{sec:largest eigenvalue}}
Suppose that the fault in state evaluation matrix $\boldsymbol{X}$ occurs at sampling time $\tau$.
The point-by-point detection is achieved by a sliding time window. The size of the window is $p\times d$, where $p$ is the number of nodes and $d$ is the sample size. The window has a step size of 1, that is, one new column of data is moved into the window and one column of historical data is removed at each time.
$d_1$ and $d_2$ are selected such that $d=d_1+d_2$, where $d_2$ is slightly larger than $p$. Thus it is to ensure that the inverse of the covariance matrix exists.

For the data $\left( \boldsymbol{x}_{k},\dots,\boldsymbol{x}_{k+d-1}\right)$ of each window, we consider the following samples from two populations:
\begin{equation}\label{Pk}
\begin{aligned}
\boldsymbol{X}_k^{(1)}=&\left( \boldsymbol{x}_{k},\dots,\boldsymbol{x}_{k+d_1-1}\right),\\ \boldsymbol{X}_k^{(2)}=&\left( \boldsymbol{x}_{k+d_1},\dots,\boldsymbol{x}_{k+d-1}\right),
\end{aligned}
\end{equation}
where $k=1,\dots,K$, $K=T-d+1$.
The normalization of each segment $\boldsymbol{X}_k^{(i)}$ is $\widetilde{\boldsymbol{X}}_k^{(i)},i=1,2,k=1,\dots,K$.
We denote the unbiased sample covariance matrices corresponding to the $\widetilde{\boldsymbol{X}}_k^{(i)}$ are $\boldsymbol{S}_k^{(i)}$
and denote the Fisher matrix as
\begin{equation}
	\boldsymbol{F}_k=\boldsymbol{S}_k^{(1)}{\boldsymbol{S}_k^{(2)}}^{-1}.\label{Fmatrix}
\end{equation}
The largest eigenvalue of $\boldsymbol{F}_k$ is defined as $\lambda_{1k}$. Then it can be judged that a fault occurs at sampling time $\tau_k=k+d-1,$ if $\lambda_{1k}$ is larger than $b$.

To decrease the random errors, we consider the occurrence of the event $\{\lambda_{1k_i}>b,1\leq k_i \leq K, i=1,\dots,s\}$, where $k_1,\dots,k_s$ are the continuous sampling time. In other words, after $s$ consecutive sampling points are judged as faults, the time of fault is finally determined by
	$$\tau=k_s+d-1.$$
The steps of DELE are as shown in Algorithm \ref{alg:1}.
\begin{algorithm}
	\caption{DELE}
	\label{alg:1}
	\begin{algorithmic}[1]	
		\REQUIRE
		The state evaluation matrix $\boldsymbol{X}=\left( \boldsymbol{x}_1,\dots,\boldsymbol{x}_T\right)$; the window width $d_1,d_2$;\\
		\ENSURE
		The time of fault $\tau_j,j=1,\dots,J$;\\
		\STATE Do Algorithm \ref{alg2} and get the shrinking intervals $\left[t_{l_j-1}+1,t_{m_j+1} \right] ,j=1,\dots,J$;\\
		\FOR{$j=1$ to $J$}
		\STATE
		 $\left(\boldsymbol{x}_{t_{l_j-1}+1},\boldsymbol{x}_{t_{m_j+1}} \right)$ replaces $\left( \boldsymbol{x}_1,\dots,\boldsymbol{x}_T\right)$
		\FOR{$k=1$ to $K$}
		\STATE
		The normalization of $\boldsymbol{X}_i^{(\ell)}: ~\widetilde{\boldsymbol{X}}_i^{(\ell)} , \ell=1,2$;\\
		\STATE Calculate the sample covariance matrices $\boldsymbol{S}_k^{(1)},\boldsymbol{S}_k^{(2)}$;\\
		\STATE $\boldsymbol{F}_k=\boldsymbol{S}_k^{(1)}{\boldsymbol{S}_k^{(2)}}^{-1}$;\\
		\STATE Calculate the largest eigenvalue of $\boldsymbol{F}_k$ be $\lambda_{1k}$;\\
		\ENDFOR	
		\IF{exists \{$\lambda_{1k_i}>b,i=1,\dots,s$\}}
		\STATE
		the time of fault: $k_s+d-1$;\\
		\ELSE{}
		\STATE
		No fault exists;\\
		\ENDIF
		\RETURN $\tau_j=k_s+d-1.$
		\ENDFOR
	\end{algorithmic}		
\end{algorithm}

\begin{figure}[htbp]
 	\centering
 	\begin{minipage}[b]{.8\linewidth}
 			\centerline{
 			\includegraphics[scale=0.4]{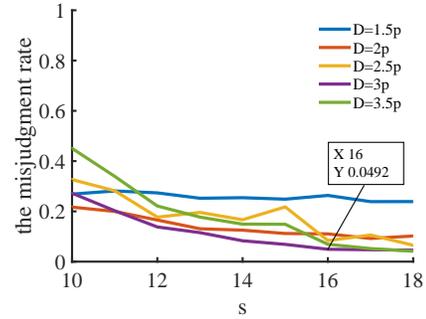}}	
 		\end{minipage}
 	\caption{The misjudgment rate under different widths $D$ and $s$.}\label{fig:s-D}
 \end{figure}

To further optimize the algorithm, the optimal $D$ and $s$ are selected through the experiment of the misjudgment rate. The value of $D$ and $s$ are selected based on the distribution and transmission networks model. To facilitate the calculation, we select the $3000th-7000th$ sampling points of the data of abnormal operation in Section \ref{sec:Adaptability}. With 5000 replications at a significance level $\alpha=0.01$, the misjudgment rate under different widths $D$ and $s$ as shown in Fig. \ref{fig:s-D}. The misjudgment rate represents the ratio that which the power grid is in normal operation but judged as abnormal. That is, the ratio of the number of experiments that the detected fault before the $2000th$ sampling point to the total number of experiments.

The result indicates the misjudgment rate is controlled effectively with the increase of $s$. In the case of $s=16, D=3p$, the random errors can be effectively decreased, and the misjudgment rate is below 0.05. When $s$ is less than 16, the misjudgment rate of the fault detection is higher. When the value of $s$ is too large, the delay time of fault detection will be increased in practical applications, resulting in a decrease in accuracy. For this distribution network model with 80 nodes ($p=80$), $s =16$, $D=3p$ are selected, that is, only after 16 consecutive sampling points are judged as faults can the occurrence of fault be finally determined.

Although the largest eigenvalue of the Fisher matrix can be well used to determine the time of fault point-by-point, it involves the calculation of eigenvalues of a  large-dimensional matrix. Therefore, to avoid the high complexity calculation, a simpler method based on the hypothesis testing method will be used to do point-by-point detection through each shrinking interval in the next part.

\subsection{DEHT: Determination by Hypotheses Testing\label{sec:test statistic}}
This subsection combines the sliding time window with the hypothesis testing method in Section \ref{sec:2.2} to determine the time of faults.
The state evaluation matrix $\boldsymbol{X}$ is divided into the segments $\boldsymbol{X}_k^{(i)},i=1,2,k=1,\dots,K$ like (\ref{Pk}).
For each segment, $\widetilde{\boldsymbol{X}}_k^{(1)}$ and $\widetilde{\boldsymbol{X}}_k^{(2)}$ are taken as samples from two populations, the covariance matrices of the two population are defined as $\boldsymbol{\Sigma}_k^{(1)}$ and $\boldsymbol{\Sigma}_k^{(2)}$, respectively.
Fault detection like the test (\ref{HOinv}) is performed at each sampling time $k+d_1-1$:
\begin{equation}
	\mathcal{H}_{0k}:\boldsymbol{\Sigma}_k^{(1)}{\boldsymbol{\Sigma}_k^{(2)}}^{-1}=\boldsymbol{I}\quad v.s.\quad \mathcal{H}_{1k}:\boldsymbol{\Sigma}_k^{(1)}{\boldsymbol{\Sigma}_k^{(2)}}^{-1}\neq \boldsymbol{I}, \label{H0k}
\end{equation}
where $k=1,\dots,K$.
If $\mathcal{H}_{0k}$ is rejected at sampling time $k+d_1-1$, that is, $L_k\in W_\alpha$, it can judge that a fault occurs at sampling time $\tau_k=k+d-1$. Here $L_k$ is the analogue of $L$ in (\ref{Lstat}), except that $\boldsymbol{F}$ is replaced by $\boldsymbol{F}_k$ defined in (\ref{Fmatrix}).

In the same way, to decrease the random errors, if the null hypothesis is rejected at $s$ consecutive sampling points, that is, $\{L_{k_i}\in W_\alpha,1\leq k_i \leq K, i=1,\dots,s\}$, where $k_1,\dots,k_s$ are the continuous sampling time. Then the time of fault can be determined by
	$$\tau=k_s+d-1.$$
The steps of DEHT are as shown in Algorithm \ref{alg:2}.
\begin{algorithm}
	\caption{DEHT}
	\label{alg:2}
	\begin{algorithmic}[1]	
	\REQUIRE
		The state evaluation matrix $\boldsymbol{X}=\left( \boldsymbol{x}_1,\dots,\boldsymbol{x}_T\right)$;	the window width $d_1,d_2$;\\
		\ENSURE
		The time of fault $\tau_j,j=1,\dots,J$;\\
		\STATE Do Algorithm \ref{alg2} and get the shrinking intervals $\left[t_{l_j-1}+1,t_{m_j+1} \right] ,j=1,\dots,J$;\\
		\FOR{$j=1$ to $J$}
		\STATE $\left(\boldsymbol{x}_{t_{l_j-1}+1},\boldsymbol{x}_{t_{m_j+1}} \right)$ replaces $\left( \boldsymbol{x}_1,\dots,\boldsymbol{x}_T\right)$
		\FOR{$k=1$ to $K$}
		\STATE
		Repeat Steps 5-7 of Algorithm \ref{alg:1} ;\\
		\STATE  $L_k=\nu_g^{-1/2}\left\lbrace  \mbox{tr}\left\lbrace (\boldsymbol{F}_k- \boldsymbol{I})^2\right\rbrace - p F_{y_\tau, y_T}(g)- \mu_g\right\rbrace$;\\
		\ENDFOR	
		\IF{exists \{$L_{k_i}\in W_\alpha,i=1,\dots,s$\}}
		\STATE
		the time of fault: $k_s+d-1$;\\
		\ELSE{}
		\STATE
		No fault exists;\\
		\ENDIF
		\RETURN $\tau_j=k_s+d-1.$
		\ENDFOR
		
	\end{algorithmic}		
\end{algorithm}

\section{Case Studies\label{sec:4}}
The effectiveness of DELE and DEHT methods is validated by the simulated data from the distribution and transmission networks, respectively. Three cases from two networks are designed as follows:

Case 1-2 consists of the data from a distribution network with 80 nodes in Section \ref{sec:Adaptability}. With a
sampling frequency of 20,000 Hz, the simulation time is 0.75s, grounding faults all start at 0.25s and are cut off at 0.45s.
To facilitate the calculation, the data of $1st-8000th$ sampling points are intercepted for simulation validation.

Case 1: Single-phase grounding by small resistance, node 66 is grounded by 10$\Omega$.

Case 2: Two nodes grounded simultaneously, and single-phase grounding faults are set at node 4 and node 11 at the same time, by 100$\Omega$ and 200$\Omega$, respectively.

Based on the Power System Analysis Software Package (PSASP), the data of Case 3 are from the IEEE 39-bus system of the transmission network. Its sampling frequency is 100Hz, and the simulation time is 5s.

Case 3: From 0 to 1.5s, normal load; from 1.5 to 2s, all loads increase to 105\%; from 2 to 5s, all loads remain in 105\%.

In order to simulate the measurement error and data fluctuation in practice, white noise $0.0001\boldsymbol{\varepsilon}$ is added to the simulation data Case 1-3, where $\boldsymbol{\varepsilon}$ follows the standard Gaussian distribution.

\subsection{The Performance of DELE in Algorithm \ref{alg:1}}
Assume that $d_1=p-10, d_2=p+10$, the sliding time window has a size of $p\times2p$ and a sliding step of 1.

The first step is to use Algorithm \ref{alg2} to screen the shrinking intervals of fault. For Case 1-2, the data is divided into 33 segments by choosing a segmentation width of $D=240$, and then 32 tests are performed. In both cases, the shrinking interval of the fault is [4560,6240].
For Case 3, similar to the selection method of distribution network, the optimal $D=58$ and $s=9$ are selected, and the screened shrinking interval is [58,500].

\begin{figure*}[htbp]
 	\centering
 	\subfigure[]
 	{
 	\begin{minipage}[b]{.3\linewidth}
 			\centerline{
 			\includegraphics[scale=0.37]{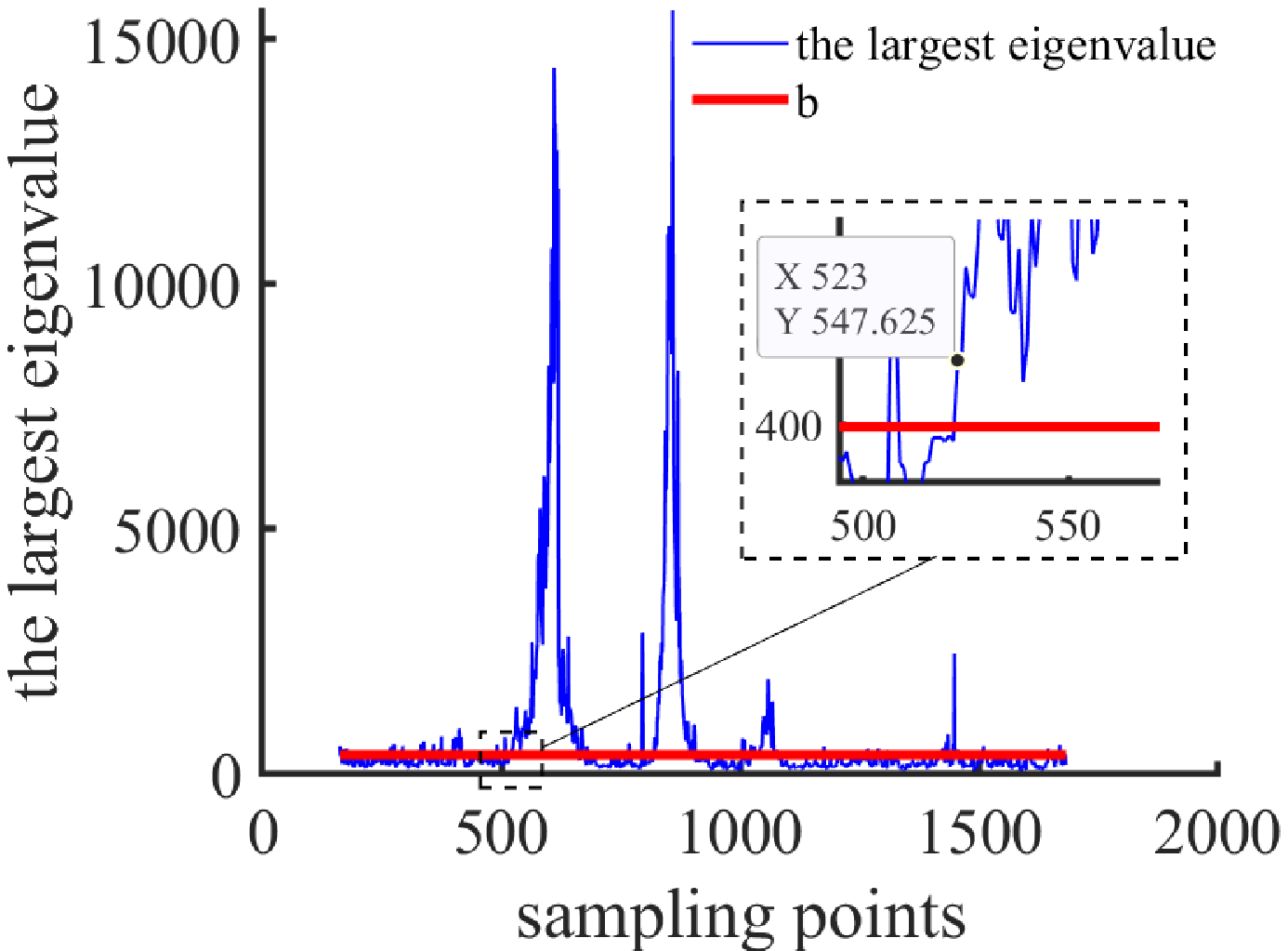}}	
 		\end{minipage}
 	}
 \subfigure[]
 	{
 	\begin{minipage}[b]{.3\linewidth}
 			\centerline{
 			\includegraphics[scale=0.37]{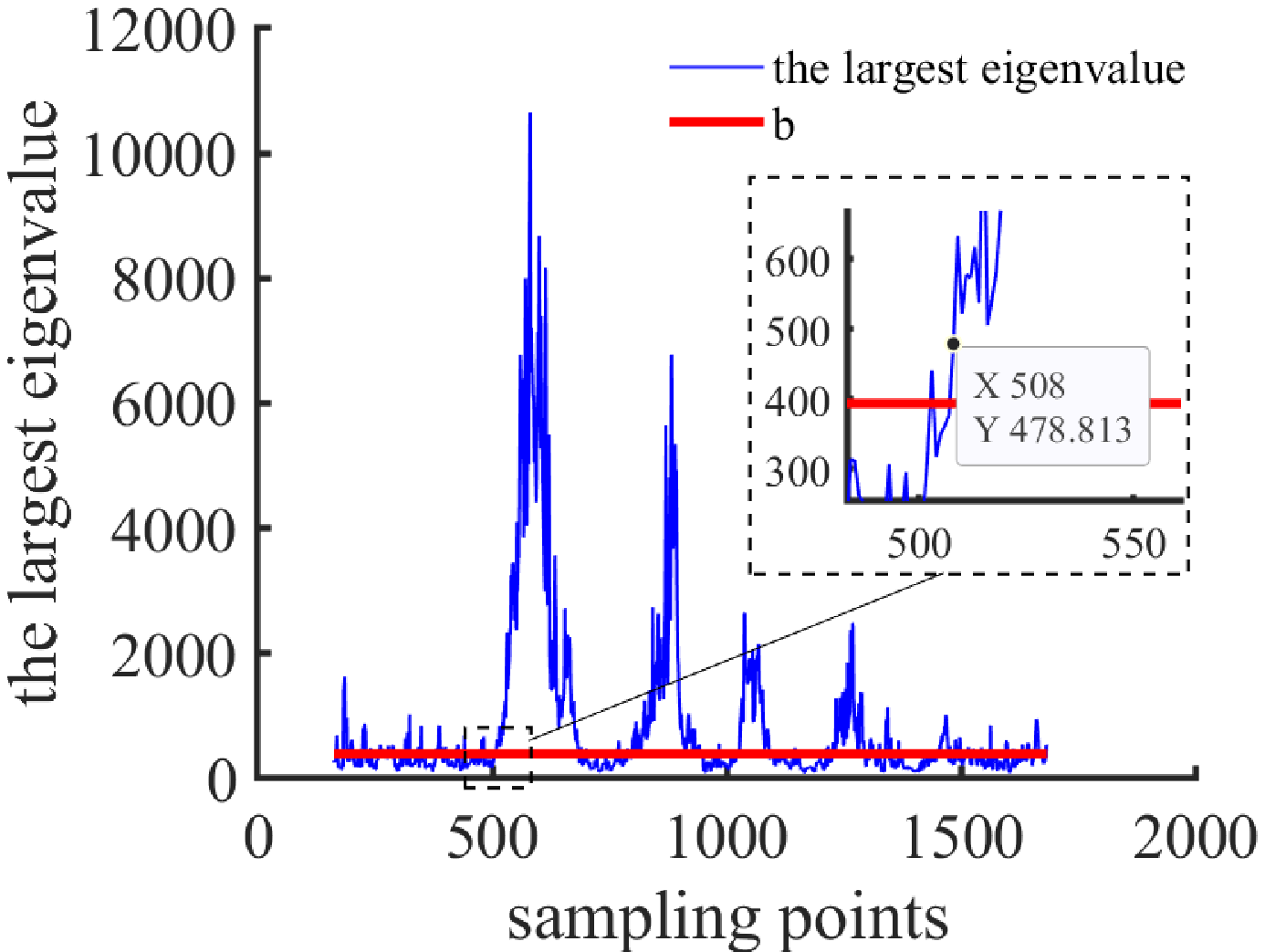}}	
 		\end{minipage}
 	}
 	 \subfigure[]
 	{
 	\begin{minipage}[b]{.3\linewidth}
 			\centerline{
 			\includegraphics[scale=0.37]{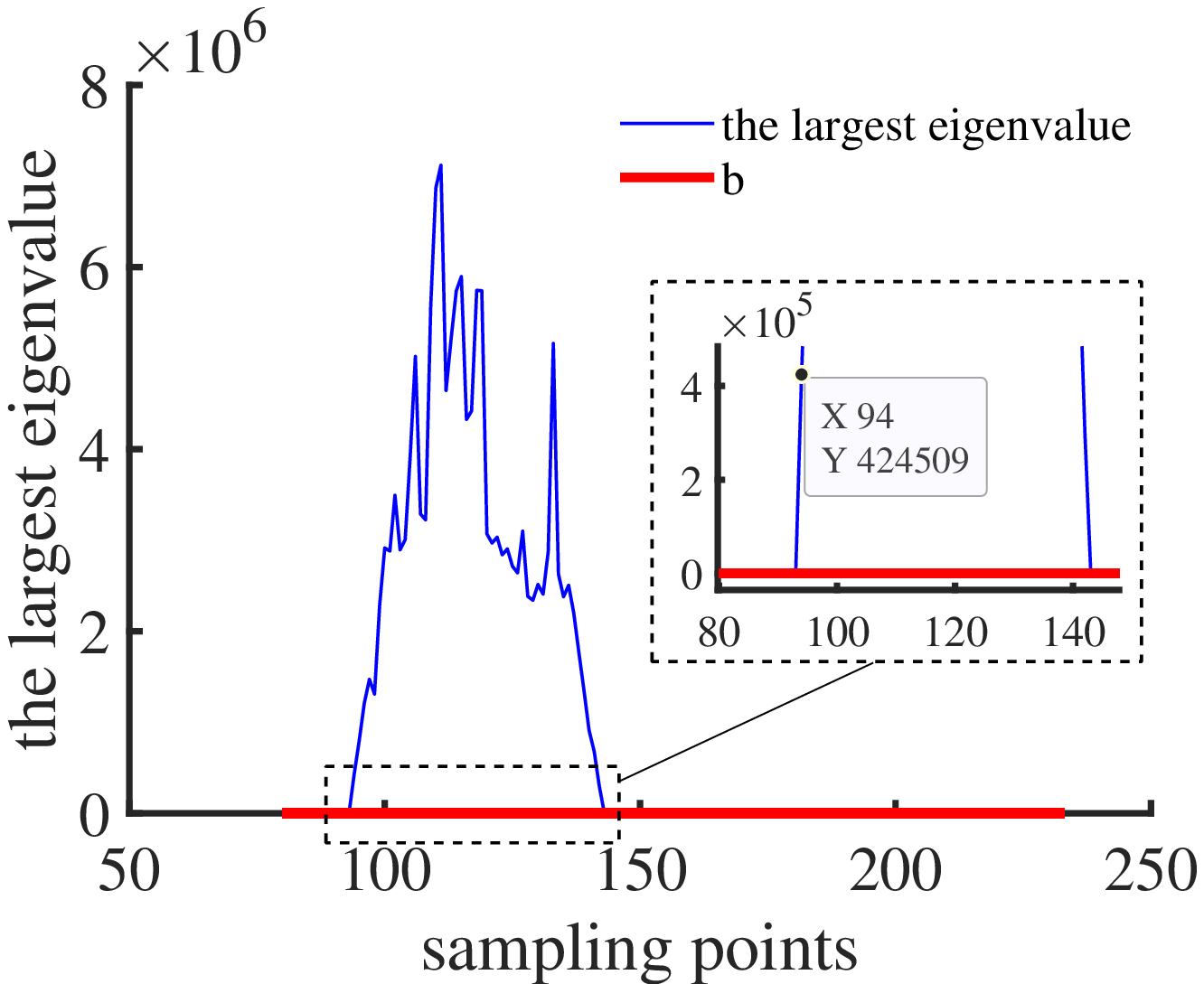}}	
 		\end{minipage}
 	}
 	\caption{The detection results of DELE within each shrinking interval. (a) The result in the interval [4560,6240] for Case 1. (b) The result in the interval [4560,6240] for Case 2. (c) The result in the interval [58,500] for Case 3.}\label{fig:methods1}
 \end{figure*}

Next, the method based on the largest eigenvalue is used to determine the time of fault within each screened shrinking interval.
The results of fault detection by DELE method are shown in Fig. \ref{fig:methods1}.
When the power grid is in normal operation, most values of $\lambda_{1k}$'s are below the red line ($y_{axis} =b$) in the figure and   few points are above the red line. When the curve of the largest eigenvalues exceeds the red line in the figure at 16 consecutive sampling points, it can be determined that the system is in an abnormal operation.

According to Fig. \ref{fig:methods1}(a), in the interval [4560,6240], from the $523rd$ sampling point, the curve of $\lambda_{1k}$'s shows a rapid upward trend and dominates the red line at more than 16 consecutive sampling points. It can be determined that there is a disturbance in the system at this time, causing the system to operate abnormally.
Therefore, the fault of the total data is identified at the $5099th$ sampling point.
In the simulation, the time of fault is set as the $5000th$ sampling point, so there is a delay of 4.95ms.
For Case 2, the detection result in the interval [4560,6240] is shown in Fig. \ref{fig:methods1}(b).
It can be seen that after the $508th$ sampling point, there are 16 consecutive points that remain above the red line, so it can be judged that the $524th$ sampling point is fault. Therefore, the fault of the total data is at the $5084th$ sampling point with a delay of 4.20ms.
For Case 3, assume that the fault starts at the $150th$ sampling point, the detected fault is the $103rd$ sampling point in the interval [58,500], with a delay of 0.11s.

\subsection{The Performance of DEHT in Algorithm \ref{alg:2}}
\begin{figure*}[htb]
	\centering	
	\subfigure[]
	{
		\begin{minipage}[b]{.3\linewidth}
			\centerline{
			\includegraphics[scale=0.37]{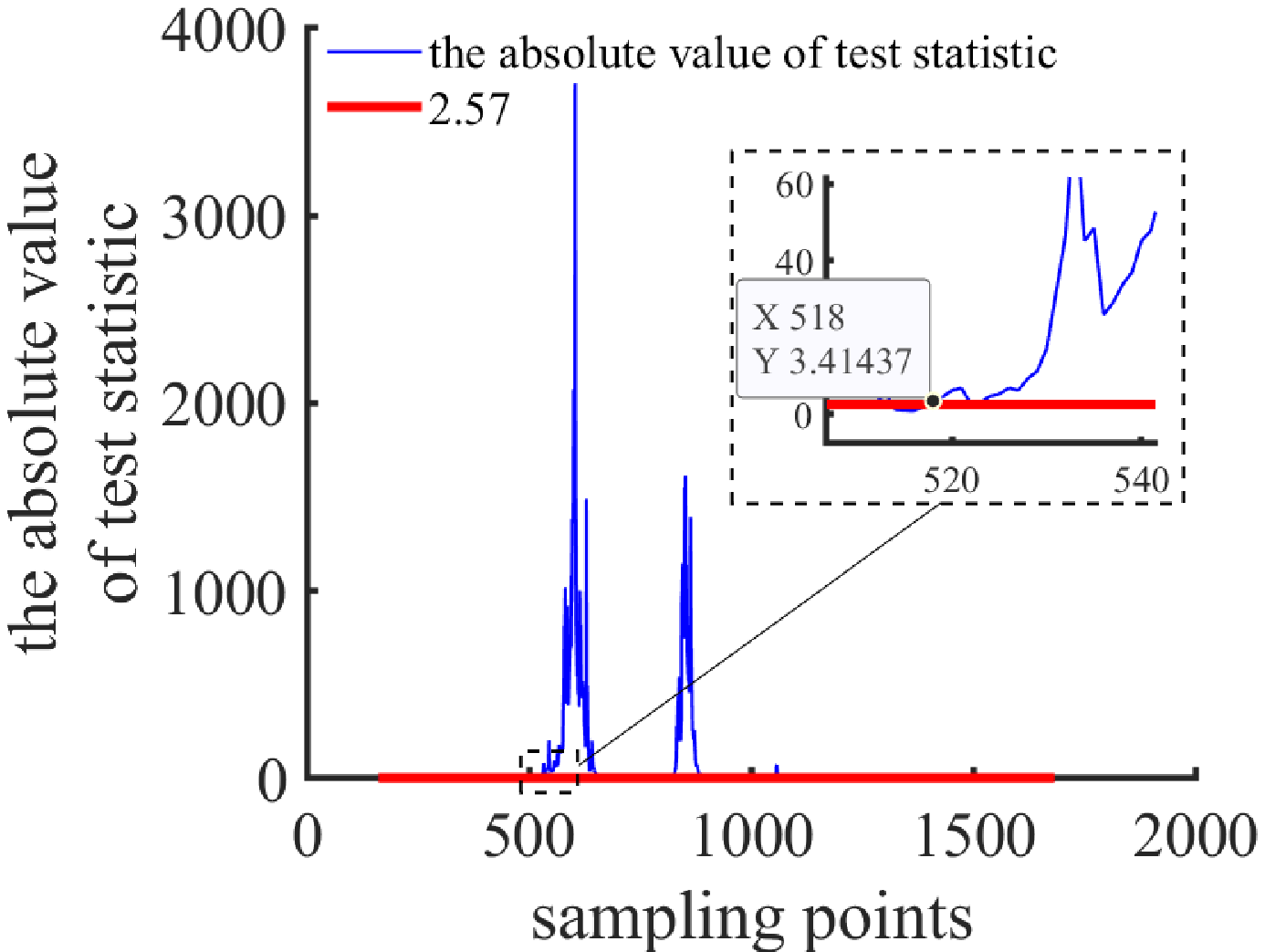}}	
		\end{minipage}
	}
	\subfigure[]
	{
		\begin{minipage}[b]{.3\linewidth}
			\centerline{
			\includegraphics[scale=0.37]{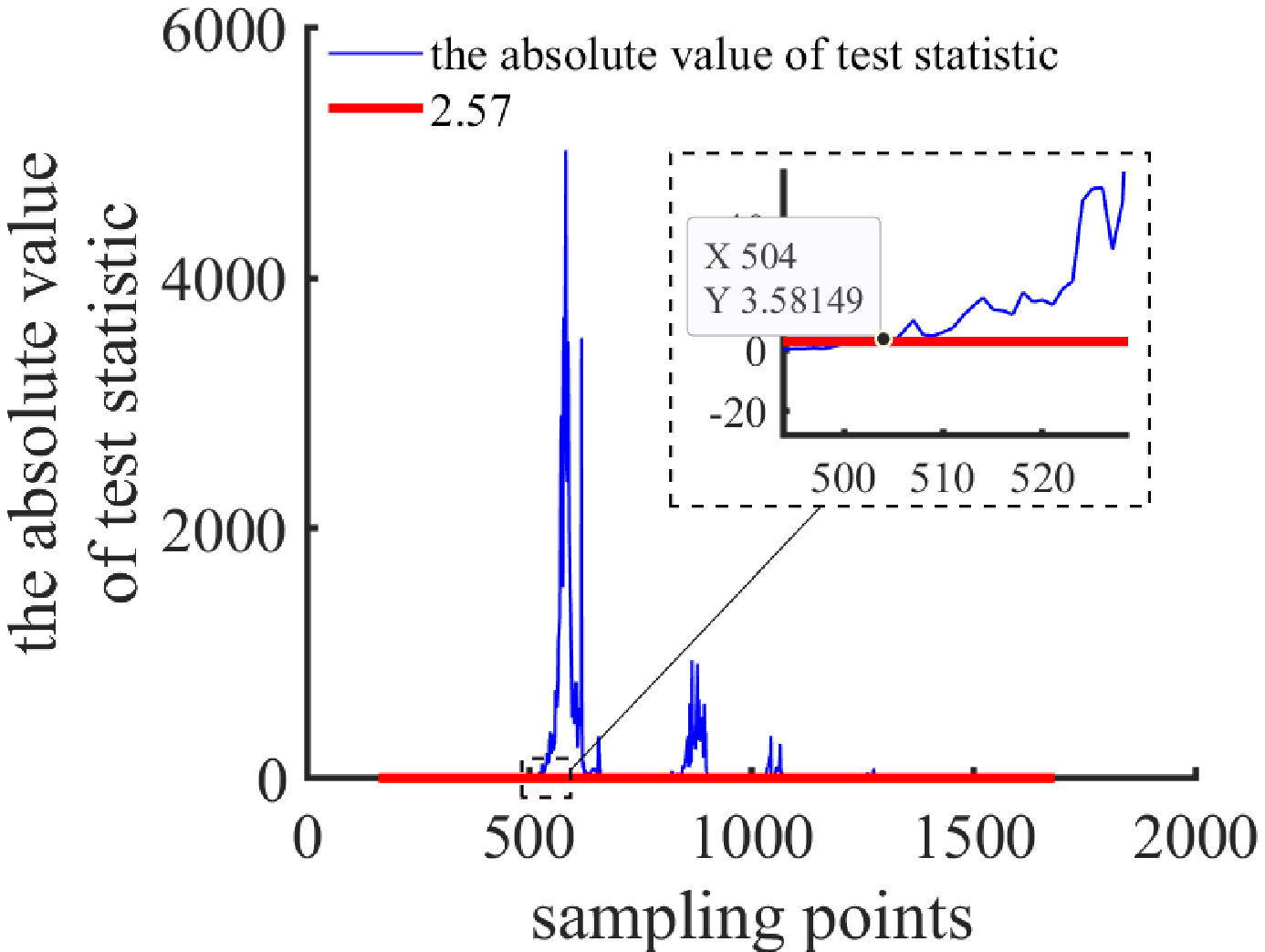}}	
		\end{minipage}
	}
	\subfigure[]
	{
		\begin{minipage}[b]{.3\linewidth}
			\centerline{
			\includegraphics[scale=0.37]{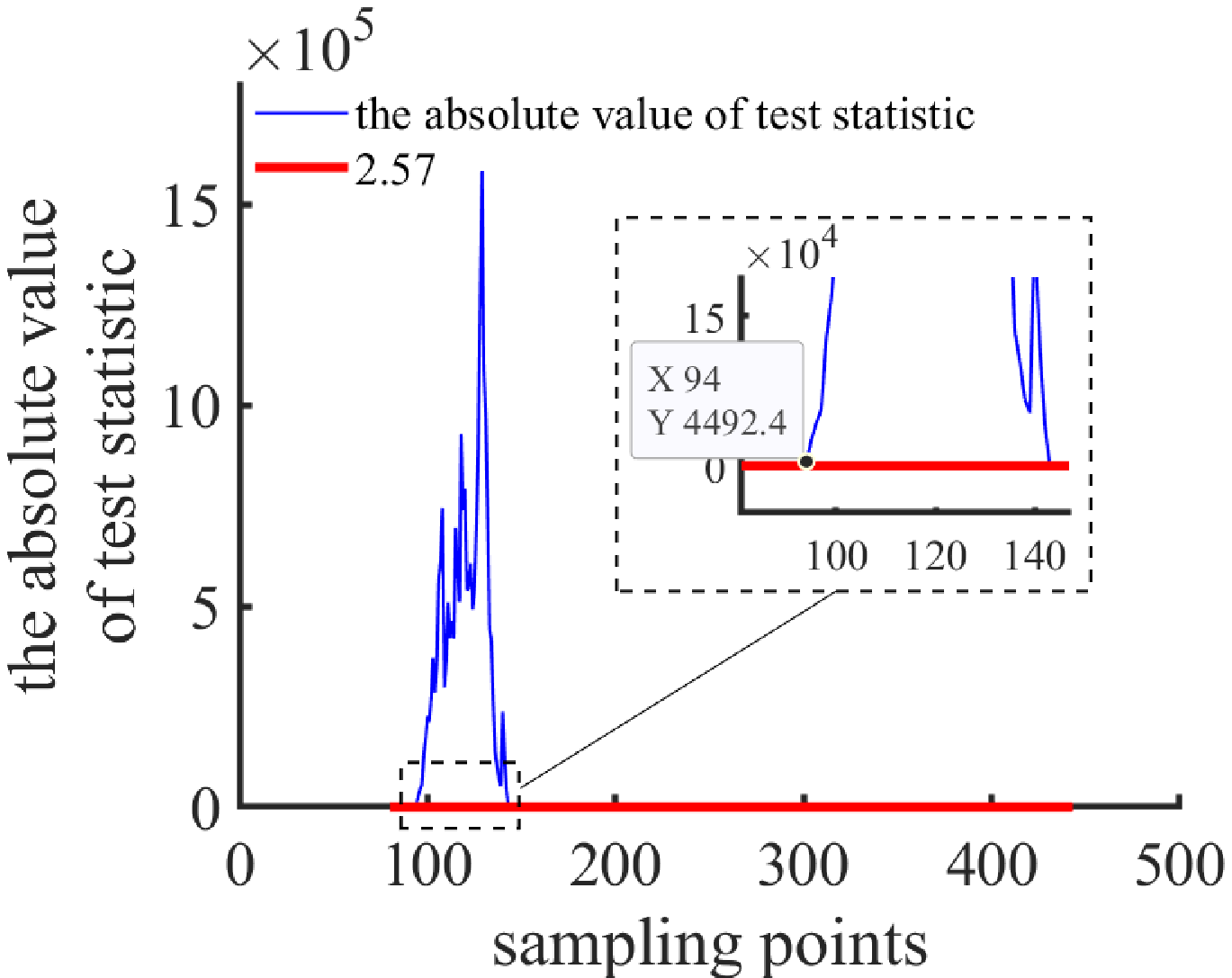}}	
		\end{minipage}
	}
	\caption{The detection results of DEHT within each shrinking interval. (a) The result in the interval [4560,6240] for Case 1. (b) The result in the interval [4560,6240] for Case 2. (c) The result in the interval [58,500] for Case 3.}\label{fig:methods2}
\end{figure*}

The data of Case 1-3 are used for the validation of DEHT method, and the obtained shrinking interval is unchanged. At a significance level $\alpha=0.01$,
Fig. \ref{fig:methods2} shows a comparison between the absolute value of test statistic $L_k$ at each sampling point and its corresponding criterion $U_{1-\alpha/2}=2.57$, which is the $1-\alpha/2$ th quantile of the standard Gaussian distribution.

For Case 1, it can been seen from Fig.  \ref{fig:methods2}(a), after the $518th$ sampling point, the situation $\{|L_k|>U_{1-\alpha/2}\}$ happens at least 16 times in a row, so it can be judged that the system is abnormal at the $5094th$ sampling point, which is 4.70ms later than the real time.
For Case 2, after the $504th$ sampling point, there are 16 consecutive points that reject the null hypothesis. Then the fault is detected at the $520th$ sampling point in the interval [4560,6240], as it shown in the Fig. \ref{fig:methods2}(b). In other words, the fault is detected at the $5080th$ sampling point with a delay of 4.00ms. For Case 3, it follows the same result as the one by DELE method, the detected time of fault is also delayed by 0.11s.

From the above analysis, DELE and DEHT methods can accurately identify the abnormal operation of the system and provide information on the time of fault. The screening steps based on the hypothesis testing method can quickly shrink the range of faults to cost less computational time. The two methods can detect faults point-by-point between each shrinking interval to achieve a higher degree of accuracy.
In addition, delays are unavoidable because the system needs a certain reaction time after a fault occurs.

\subsection{Comparisons and Evaluations\label{sec:4.3}}
A comparison and evaluation of the proposed methods with other methods are given in this section.
We focus on the accuracy and rapidity of DELE and DEHT methods compared to other existing fault detection methods: the method based on M-P law in \cite{17}, the method based on MSR in \cite{MSR}, and the method based on SVM in \cite{SVM}.
Reference \cite{17} screens the eigenvalues of sample covariance matrix larger than the right endpoint of the supporting set of the M-P law as abnormal eigenvalues. On this basis, the M-P method in our paper determines that a fault occurs after $s$ consecutive sampling points are abnormal eigenvalues. 
Reference \cite{MSR} determines the faults by judging whether the MSR drops dramatically and is smaller than the inner ring radius of the ring law, the faults are determined by the MSR of $s$ consecutive sampling points is smaller than the inner ring radius in our paper.
In \cite{SVM}, SVM is used to classify power quality disturbance signals. The SVM method in our paper classifies normal and abnormal operations and determines the occurrence of faults after $s$ consecutive sampling points are classified into abnormal classes.

\begin{figure*}[htbp]
\centering
		\begin{minipage}[b]{.3\linewidth}
			\centerline{
		\includegraphics[scale=0.35]{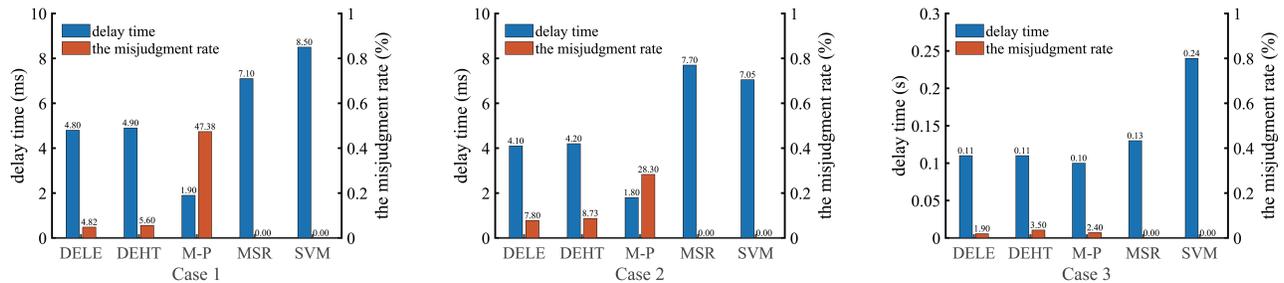}}
		\end{minipage}
	\caption{Accuracy: Comparisons of five methods on delay time and the misjudgment rate.}\label{fig:accuracy}
\end{figure*}

\begin{figure}[htbp]
	\centering
	\begin{minipage}[b]{.8\linewidth}
		\centerline{
			\includegraphics[scale=0.29]{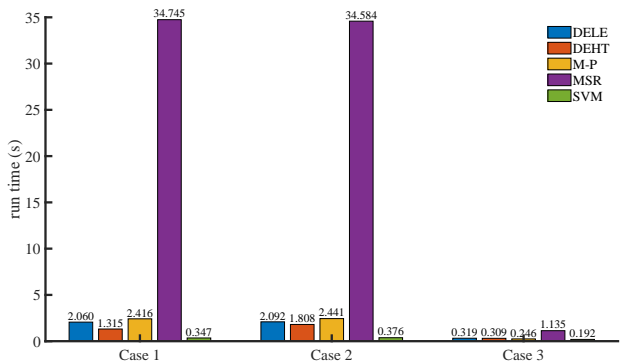}}	
	\end{minipage}
	\caption{Rapidity: Comparisons of five methods on run time.}\label{fig:time}
\end{figure}

To validate the accuracy of the two proposed methods for the distribution and transmission networks, the DELE and DEHT methods are compared with M-P, MSR, and SVM methods on delay time and misjudgment rates of fault detection.
Under the condition of $s=16$ (Case 1-2) and $s=9$ (Case 3), Case 1-3 are detected by five methods with 1000 replications, and their (average) delay time and misjudgment rates are calculated. Figure. \ref{fig:accuracy} shows the delay time and misjudgment rate of the compared methods.
To compare the time consumed by different methods, Case 1-3 are detected by five methods, respectively, and Fig. \ref{fig:time} shows the histogram of the run time of the comparative five methods.

From Fig. \ref{fig:accuracy}, the DELE, DEHT, and M-P methods have a very short delay time. In contrast, the delay time of MSR and SVM methods is large. For Case 1-2, the value of the misjudgment rate of DELE, MSR, and SVM methods is less than 5\%, which means they have a good performance in terms of misjudgment rate. However, the accuracy of MSR and SVM methods is still low due to their long delay time. The M-P method has a high misjudgment rate, although it performs well in terms of delay time. To achieve a misjudgment rate close to the level of other methods, a larger value of $s$ is required, which leads to a large increase in delay time and results in a decrease in the accuracy of the M-P method. For Case 3, DELE, DEHT, and M-P methods have high accuracy with little difference, and the accuracy is much better than MSR and SVM methods. In conclusion, the DELE and DEHT methods have a high degree of accuracy in terms of both delay time and misjudgment rate, especially DELE method has the best performance in detection accuracy.

Comparing the time consumed by the five methods, firstly, although the SVM method takes the least amount of time, it consumes a lot of machine memory and computing time during the modeling process. Moreover, for different test sets, the classification results of SVM method are quite different, which is not conducive to the stability of the data.
Secondly, for Case 1-2, the proposed DELE and DEHT methods are faster than M-P and MSR methods. In particular, the DEHT method greatly improves the speed of fault detection among these four methods, as other methods involve a large number of calculations of eigenvalues. For Case 3, the detection time of DELE and DEHT methods is slightly slower than M-P method but much faster than MSR method.
Overall, the DELE and DEHT methods make the identification of fault more rapid through the interval screening method,  especially in large samples and dimensions situations.

\section{Conclusions\label{sec:5}}
Based on the theory of the random Fisher matrix, the high dimensional state evaluation matrix from the PMUs-based wide-area measurement system is analyzed.
Firstly, this paper constructs the CLT for the LSS of the Fisher matrix and proposes the interval screening method for the faults on the basis of this CLT.
Secondly, two point-by-point methods are also provided to detect the state of power grids: one is the detection by the largest eigenvalue of the Fisher matrix, and the other is the detection by the statistic proposed. Furthermore, in order to reduce the cost of the complex computation, the two point-by-point detection methods and the interval screening method are combined respectively to obtain two fast determination methods for the faults: DELE and DEHT.

The newly proposed fast determination methods have several advantages: First, there is no need to model the system, which improves the deficiency of modeling analysis in abnormal state detection of power systems; Second, combining the fault interval screening method, the proposed fault detection methods perform a great improvement in the recognition speed; Moreover, the analysis of the state evaluation matrix based on the Fisher random matrix can timely and effectively reflect the data changes when faults occur in the operation of power grids. Therefore, the new methods have high detection accuracy.

In the future, we may introduce the limiting theory of the general large-dimensional correlation coefficient matrix into this field, thus the state evaluation matrix can be considered without the constraint of the independent assumption.

\section*{Acknowledgment}
The authors are grateful to the Editor, the Associate Editors, and referees for their review of the paper.

\section*{Supplementary Material}
The Supplementary Material contains the performance of the proposed test statistic.

\noindent\textbf{Ke Chen}  received the B.S. degree from China University of Mining and Technology-Beijing, Beijing, China, in 2021. She is currently studying at the School of Mathematics and Statistics at Xi'an Jiaotong University in Xi'an, China. Her research interests include random matrix theory.\\

\noindent\textbf{Dandan Jiang} is a Professor with the School of Mathematics and Statistics at Xi'an Jiaotong University in Xi'an, China.
Her research interests include random matrix theory, statistical inference of high-dimensional data and its application in wireless communication.
\\

\noindent\textbf{Bo Wang} received the Ph.D. degree in Computer Science from Wuhan University, Wuhan, China, in 2006, and performed postdoctoral research in Electrical Engineering from Wuhan university from 2007 to 2009. He is currently a Professor with the School of Electrical Engineering, Wuhan University, Wuhan, China. His research interests include power system online assessment, big data, and integrated energy systems. \\

\noindent\textbf{Hongxia Wang} received the M.Sc. degree  from Wuhan University, Wuhan, China, in 2020. She is currently pursuing the Ph.D. degree in Electrical Engineering from Wuhan University, Wuhan, China. Her research interests include big data and data fusion in power systems. \\

\end{document}